\documentclass[aps,prl,twocolumn,prabib,amsmath,amssymb,superscriptaddress,notitlepage]{revtex4-1}

\usepackage{graphicx}
\usepackage{bm}
\usepackage{dcolumn}
\usepackage{natbib}
\usepackage{epstopdf}
\usepackage{graphicx}
\usepackage{epsfig}
\usepackage[pdfstartview=FitH]{hyperref}
\usepackage{color}
\usepackage{ulem}

\newcommand{\rr}{{\bf r}}

\begin{document}
\title{Tuning topological phase and quantum anomalous Hall effect  by interaction in quadratic band touching systems}

\author{Tian-Sheng Zeng}
\affiliation{Department of Physics, The University of Texas at Dallas, Richardson, Texas 75080, USA}
\author{W. Zhu}
\affiliation{Theoretical Division, T-4 and CNLS, Los Alamos National Laboratory, Los Alamos, New Mexico 87545, USA}
\affiliation{Westlake Institute for Advanced Study, Hangzhou, 300024, P. R. China}
\email{phwzhu@gmail.com}
\author{D. N. Sheng}
\affiliation{Department of Physics and Astronomy, California State University, Northridge, California 91330, USA}
\email{donna.sheng1@csun.edu}
\date{\today}

\begin{abstract}
Interaction driven topological phases significantly enrich the class of topological materials and thus are of great importance.
Here, we study the phase diagram of interacting spinless fermions filling the two-dimensional checkerboard lattice with a quadratic band touching (QBT) point.
By developing new diagnosis based on the state-of-the-art density-matrix renormalization group and exact diagonalization,
we determine accurate quantum phase diagram for such a system at half-filling
with  three distinct phases.  For  weak  nearest-neighboring interactions,
we demonstrate the instability of the QBT towards an interaction-driven spontaneous quantum anomalous Hall (QAH) effect.
For strong interactions, the system breaks the rotational symmetry realizing  a nematic charge-density-wave (CDW) phase.
Interestingly, for intermediate interactions we discover a symmetry-broken bond-ordered critical phase sandwiched in between the QAH and CDW phases,
which splits the QBT into two Dirac points  driven by interaction.
Instead of the direct transition between QAH and CDW phases, our identification of an intermediate phase
sheds new light on the theoretical understanding of the interaction-driven phases in QBT  systems.
\end{abstract}

\maketitle

Recently topological phases of matter in the band structure with nontrivial topological Berry phase become an exciting research area of modern physics,
culminating in the experimental observations of Haldane-honeycomb insulator~\cite{Jotzu2014} and the quantum anomalous Hall (QAH) effect in topological insulator~\cite{Chang2013}.
Generally speaking,
in such kind of systems with a nontrivial topological invariant Chern number \cite{Thouless1982},
exemplified by the integer quantum Hall effect in the absence of magnetic field~\cite{Haldane1988},
time-reversal symmetry (TRS) breaking of band structure is typically necessary.
For topologically trivial band structures with zero Chern number,
it was proposed that the strong correlation between electrons can also induce spontaneous TRS breaking at mean field level~\cite{Wu2004,Raghu2008}
and lead to QAH  effect in Dirac semimetals.
These interaction-induced topological phases are interesting, since they can significantly enrich the class of topological materials~\cite{Simon2015,Zhu2016a,Kourtis2017}.
However such a mechanism  is challenged by subsequent numerical simulations~\cite{Castro2013,Daghofer,Motruk2015,Capponi2015}, which do not support the interaction driven QAH effect.
Alternatively, it has been identified that a quadratic band touching (QBT) point has topological feature, and
can be driven towards a QAH phase with TRS breaking even under arbitrary weak interactions,
while strong interactions may lead to other competing phases~\cite{Sun2009,Sun2011b}.

\begin{figure}[b]
  \includegraphics[height=1.5in,width=3.4in]{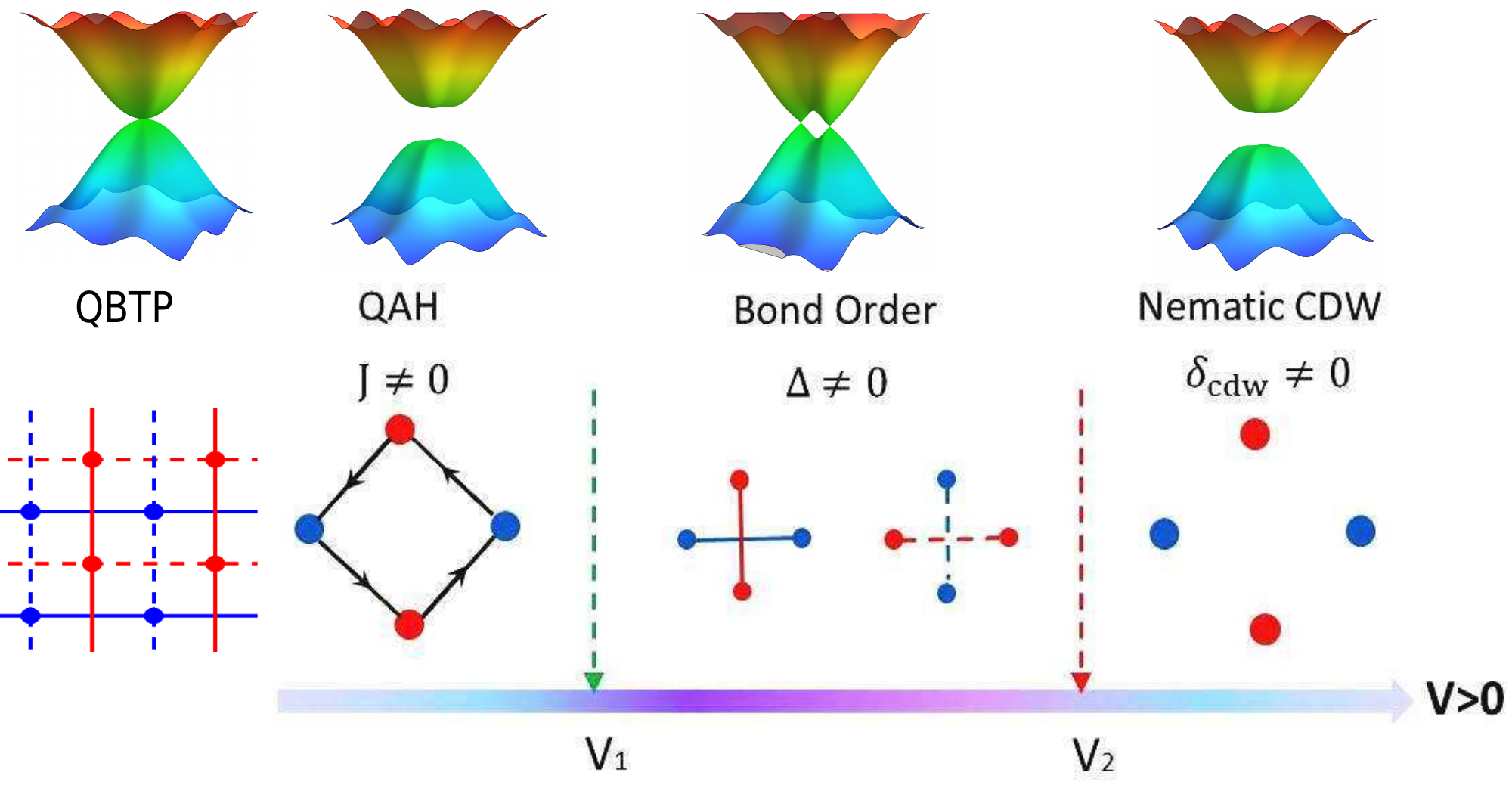}
  \caption{\label{phase}(Color online) Phase diagram of spinless fermion at half-filling in the topologically trivial checkerboard lattice with a QBT point,
  as the nearest-neighboring interaction $V$ increases. The two sublattice sites $A,B$ are labeled by blue and red colors, respectively.
  Please see the main text for the definition of order parameters.  }
\end{figure}

The theoretical prediction of such interaction-driven topological phases~\cite{Wen2010,Weeks2010,Kuruta2011,Fiete2011,Yang2011,Tsai2015,Dauphin2016,Herbut2016,Chen2017},
stimulates extensive studies  by more rigorously theoretical and numerical methods,
including low energy renormalization group approach in $C_4$ symmetric checkerboard lattice~\cite{Uebelacker2011,Murray2014,Herbut2014,Wang2017} and bilayer graphene~\cite{Nandkishore2010,Pujari2016},
first-principle calculations in spin-dependent optical square lattice~\cite{Kitamura2015} and halogenated hematite nanosheets~\cite{Liang2017}, and recently the unbiased
numerical exact diagonalization (ED) diagnosis in both $C_4$ symmetric checkerboard lattice~\cite{Wu2016} and $C_6$ symmetric Kagome lattice~\cite{Zhu2016b}.
However, the stability  of the QAH effect in the presence of  weak interaction has not
been settled. Based on the mean-field theory,  the gap protecting the QAH state is exponentially small
for weak interaction, making such a phase difficult to be identified.
A recent density-matrix renormalization group  (DMRG) study finds a semimetal phase for  Kagome QBT systems
with nearest neighboring interactions \cite{Nishimoto2010},
while adding further long-ranged interactions ($V_1 \sim V_2 \sim V_3$) favors a robust gapped QAH state~\cite{Zhu2016b}.
Thus, the weak interaction effect on the QBT point remains elusive,
because finite-size calculations are incapable of detecting extremely small energy gap.
Furthermore, the issue of the quantum phase transition between the interaction-driven topological phase and other phases is hardly touched.
While ED analysis points to a first-order phase transition from the QAH phase to the charge density wave phase without intermediate  phase~\cite{Wu2016,Zhu2016b,Hui2017},
the interesting scenario of intermediate Dirac liquid phase in such QBT systems
deserves to be explored by applying controlled numerical methods.

In this work,  we  develop accurate numerical diagnosis for the topological phases to address these challenge issues
 through the state-of-the-art DMRG and ED simulations.  A schematic diagram of our main results is shown in Fig.~\ref{phase}
for interacting spinless fermions occupying the topologically trivial checkerboard lattice with
a QBT.  For weak interactions $V<V_1$, by applying the Hellmann-Feynman theorem,
we demonstrate the emergence of interaction-driven QAH phase whose topological properties are featured by two degenerate TRS breaking ground states arising
from a pair of  Kramers degenerate states with opposite chiralities and integer quantized topological Hall conductances.
Remarkably, for intermediate interactions $V_1<V<V_2$ we establish the existence of a new gapless critical phase as a sublattice bond-ordered nematic phase,
while the nematic CDW phase appears  in the strongly interacting regime.
Our comprehensive DMRG and ED studies can access large system sizes  to establish the stability of the QAH state with weak interaction at the thermodynamic limit, and
illustrate the interaction controlling of the  anomalous dissipationless Hall transport properties of the QBT systems through opening and closing of the energy excitation gap.

\section*{Results}
\textit{Hamiltonian.---}
We consider the spinless fermions in the topologically trivial checkerboard lattice model,
\begin{eqnarray}
  H&=&-t\!\sum_{\langle\rr,\rr'\rangle}\!a_{\rr}^{\dag}b_{\rr'}-\!\!\sum_{\langle\langle\rr,\rr'\rangle\rangle}\!\big( t_{\rr,\rr'}^aa_{\rr}^{\dag}a_{\rr'}+t_{\rr,\rr'}^bb_{\rr}^{\dag}b_{\rr'}\big)\nonumber\\
   &&+ h.c. + V\sum_{\langle\rr,\rr'\rangle}\!n_{\rr}^an_{\rr'}^b + m\sum_{\rr}(n_{\rr}^a-n_{\rr}^b).\label{hal}
\end{eqnarray}
Here $a_{\rr},b_{\rr'}$ are the particle annihilation operators for sublattices A,B respectively,
and $n_{\rr}^a=a_{\rr}^{\dag}a_{\rr},n_{\rr}^b=b_{\rr}^{\dag}b_{\rr}$ the particle number operators at site $\rr$.
$\langle\ldots\rangle$ and $\langle\langle\ldots\rangle\rangle$ denote the nearest-neighbor and the next-nearest-neighbor pairs of sites on a checkerboard lattice, respectively.
As shown in Fig.~\ref{phase}, the hopping amplitudes $t_{\rr,\rr'}^a=t'$ for solid lines and $t_{\rr,\rr'}^a=-t'$ for dashed lines in sublattice A,
while $t_{\rr,\rr'}^b=t'$ for solid lines and $t_{\rr,\rr'}^b=-t'$ for dashed lines in sublattice B. 
The single-particle dispersion hosts a QBT  at $(k_x,k_y)=(\pi,\pi)$ with Berry flux $\pm2\pi$, protected by time-reversal symmetry and $C_4$ rotational symmetry.
Note that by adding a nonzero opposite shift  $\delta\neq0$ to the hopping $t_{\rr,\rr'}^{a(b)}=t_{\rr,\rr'}^{a(b)}+(-)\delta $   of both sublattices,  the system $H$ breaks $C_4$ symmetry down to $C_2$ symmetry, and the quadratic band touching splits into two
Dirac points 
with gapless particle-hole symmetric dispersions.

\begin{figure}[t]
  \includegraphics[height=1.5in,width=3.4in]{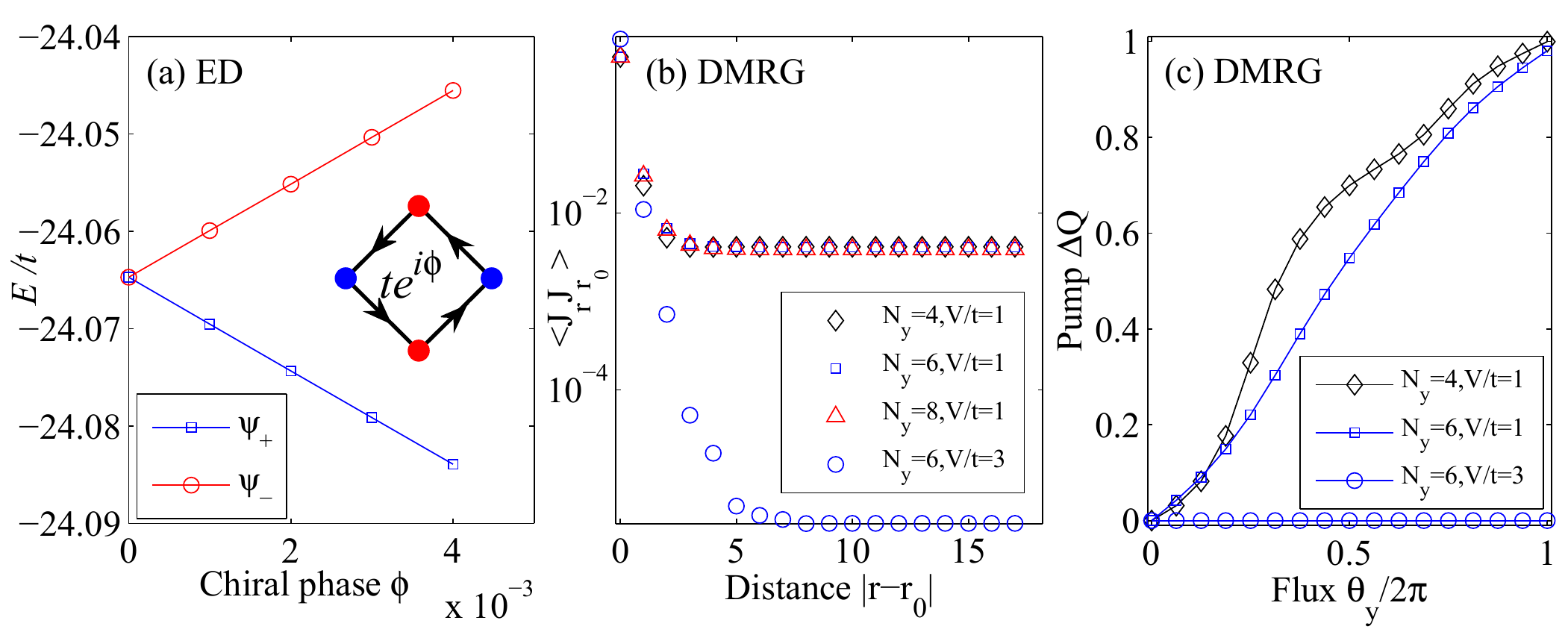}
  \caption{\label{pump}(Color online)
  Numerical diagnosis of QAH phase.
  (a) The doublet ground state energies $E_{\pm}(\phi)$, in response to the vanishingly small TRS breaking perturbation term, as a function of the phase $\phi$.
  The inset depicts the TRS breaking hopping phase in the anti-clockwise direction per plaquette between nearest-neighboring pairs.
  (b) Infinite DMRG results for current-current correlations versus distance for different cylindrical widths $L_y=2N_y$.
  (c) The adiabatic charge pumping under the insertion of flux quantum $\theta_y$. The parameters $t'=0.8t,\delta=m=0$, and the maximally kept number of states $3000$.}
\end{figure}

\textit{QAH Phase.---}
We first  address the critical issue if a robust QAH effect can be stabilized with the presence of weak and nearest interaction.
We begin with studying  the ground state properties up to a maximum system sizes $N_s=32$ based on ED.
The geometry hosts both $C_4$ point-group symmetry and time-reversal symmetry, and we  find an exact two-fold ground state degeneracy $|\psi_{\pm}\rangle$ at momentum $K=(0,0)$ below
excitation continuum for systems with finite interaction $V$.
This pair of degenerate eigenstates are also eigenstates of $C_4$ rotation with eigenvalues $\pm i$, respectively, which serves as important evidence of time-reversal
symmetry breaking as long as the energy excitation gap remains open for large systems.
In order to illustrate  two TRS spontaneously breaking states with opposite chiralities,
we consider the system response to the TRS breaking perturbation hopping phase $e^{i\phi}a_{\rr}^{\dag}b_{\rr'}+h.c.$ to  the nearest-neighbor A and  B sites
(the phase $\phi$ is a tiny detecting flux per plaquette for detecting QAH order). From the Hellmann--Feynman theorem~\cite{Hellmann,Feynman}, we can derive the TRS breaking chiral
bond current $J_{\rr}=i(a_{\rr}^{\dag}b_{\rr'}-b_{\rr'}^{\dag}a_{\rr})$ between nearest-neighboring sites from the linear response of the ground state energy, as
\begin{equation}
  \langle\psi_{\pm}|J_{\rr}|\psi_{\pm}\rangle=\frac{1}{2N_s}\frac{\partial E_{\pm}(\phi)}{\partial\phi}\Big|_{\phi=0},\label{current}
\end{equation}
where $E_{\pm}(\phi)=\langle\psi_{\pm}|H(\phi)|\psi_{\pm}\rangle$. As indicated in Fig.~\ref{pump}(a),
we can see $\langle\psi_{+}|J_{\rr}|\psi_{+}\rangle=-\langle\psi_{-}|J_{\rr}|\psi_{-}\rangle\simeq0.067$,
implying the opposite chiralities of TRS breaking for weak interactions.
To extract the topological invariants of the doublet ground states for any value $\phi$,
we utilize the twisted boundary conditions $\psi(\phi;\rr+N_{\alpha}\widehat{e}_{\alpha})=\psi(\phi;\rr)\exp(i\theta_{\alpha})$
where $\theta_{\alpha}$ is the twisted angle in the $\alpha$ (x or y)-direction. The system is periodic when one flux quantum $\theta_{\alpha}=0\rightarrow2\pi$ is inserted.
Meanwhile, the many-body Chern number of the ground state wavefunction $\psi_{\pm}(\phi)$ is defined as~\cite{Sheng2003,Sheng2006}
\begin{equation}
    C_{\pm}(\phi)=\!\!\int\!\frac{d\theta_{x}d\theta_{y}}{2\pi i} \left[\langle{\frac{\partial\psi_{\pm}}{\partial\theta_x}}|{\frac{\partial\psi_{\pm}}{\partial\theta_y}}\rangle
-\langle{\frac{\partial\psi_{\pm}}{\partial\theta_y}}|{\frac{\partial\psi_{\pm}}{\partial\theta_x}}\rangle\right].\nonumber
\end{equation}
We identify the topological invariants $C_{\pm}(\phi)=\pm1$ for the two-fold ground states $|\psi_{\pm}(\phi)\rangle$ under an infinitesimal TRS breaking phase $\phi\ll1$.
Due to the adiabatic connection between the $\psi_{\pm}(\phi)$ and $\psi_{\pm}(0)$ as shown in Fig.~\ref{pump}(a),
we can obtain the topological invariants for these doublet states $C_{\pm}=C_{\pm}(\phi\rightarrow 0)=\pm1$.
On the contrary,  for strong interactions, we find that the expectation value of $J_{\rr}$ in the ground state vanishes precisely,
and the topological invariants $C_{\pm}(\phi)=0$, signalling a topologically trivial nematic CDW phase.
Indeed,  the density structure factor $S=\frac{1}{N_s}\sum_{\alpha,\beta}\sum_{\rr,\rr'}(-1)^{\alpha}(-1)^{\beta}\langle n_{\rr}^{\alpha}n_{\rr'}^{\beta}\rangle$ shows
a strong peak (where $\alpha,\beta\in\{A,B\}$ denote sublattice indices, $(-1)^{\alpha}=1 (-1)$ for $\alpha=A (B)$) for
such a CDW phase.

To access  larger system sizes to establish the stability of the QAH, we exploit an unbiased DMRG approach using a cylindrical geometry up to a maximum width $L_y=16$ ($N_y=8$). By randomly choosing different initial states in DMRG simulations, we can obtain two different ground states $|\psi_{\pm}\rangle$ with degenerate energies
$E_{+}\simeq E_{-}$ and opposite chiral circulating loop currents per plaquette $\langle\psi_{\pm}|J_{\rr}|\psi_{\pm}\rangle$, from a DMRG algorithm allowing complex wavefunctions.
In Fig.~\ref{pump}(b), we measure the current-current correlation functions $\langle\psi_{+}|J_{\rr}J_{\rr_0}|\psi_{+}\rangle$ between nearest-neighboring bonds $\langle \rr,\rr'\rangle$ and $\langle \rr_0,\rr_{0}'\rangle$, with the distance $|\rr-\rr_0|$. For different system sizes, the bond current long-range order parameter $\langle\psi_{+}|J_{\rr}|\psi_{+}\rangle=\lim\sqrt{\langle\psi_{+}|J_{\rr}J_{\rr_0}|\psi_{+}\rangle}$ for weak interactions persists to a finite value at the large distance $|\rr-\rr_{0}|$ limit. Meanwhile in Fig.~\ref{pump}(b), we also characterize the topological nature of the ground state from its topological charge pumping by inserting one U(1) charge flux quantum $\theta_{y}=\theta$ from $\theta=0$ to $\theta=2\pi$ in the periodic $y$-direction of the cylinder system based on the  adiabatic DMRG~\cite{Gong2014} in connection to the quantized Hall conductance.
Here we partition the lattice system on the cylinder along the $x$-direction into two halves with equal lattice sites. The transverse transfer of the total charge from the right side to the left side in the $x$-direction is encoded by the expectation value $Q(\theta)=tr[\widehat{\rho}_L(\theta)\widehat{N}_{L}]$. $N_{L}$ is the particle number in the left cylinder part, and $\widehat{\rho}_L$ the reduced density matrix of the corresponding left part. Under the inserting of the flux $\theta_{y}=\theta$ in the $y$-direction, the change of
$Q(\theta)$ indicates the transverse charge transfer from the right side to the left side in the $x$-direction, induced by the
topological Hall conductances of the state  $|\psi_{\pm}\rangle$. From Fig.~\ref{pump}(c), we obtain a nearly quantized transverse Hall conductance $C_{\pm}=\Delta Q=Q(2\pi)-Q(0)=\pm1$
for these two degenerate ground states $|\psi_{\pm}\rangle$.
Again, for strong interactions, the current-current correlation functions are vanishingly small, and the charge pumping disappears, signaling the absence of a QAH phase.
Significantly,  our results  establish that any weak interaction would drive the system into the QAH phase.

\begin{figure}[t]
  \includegraphics[height=1.9in,width=3.4in]{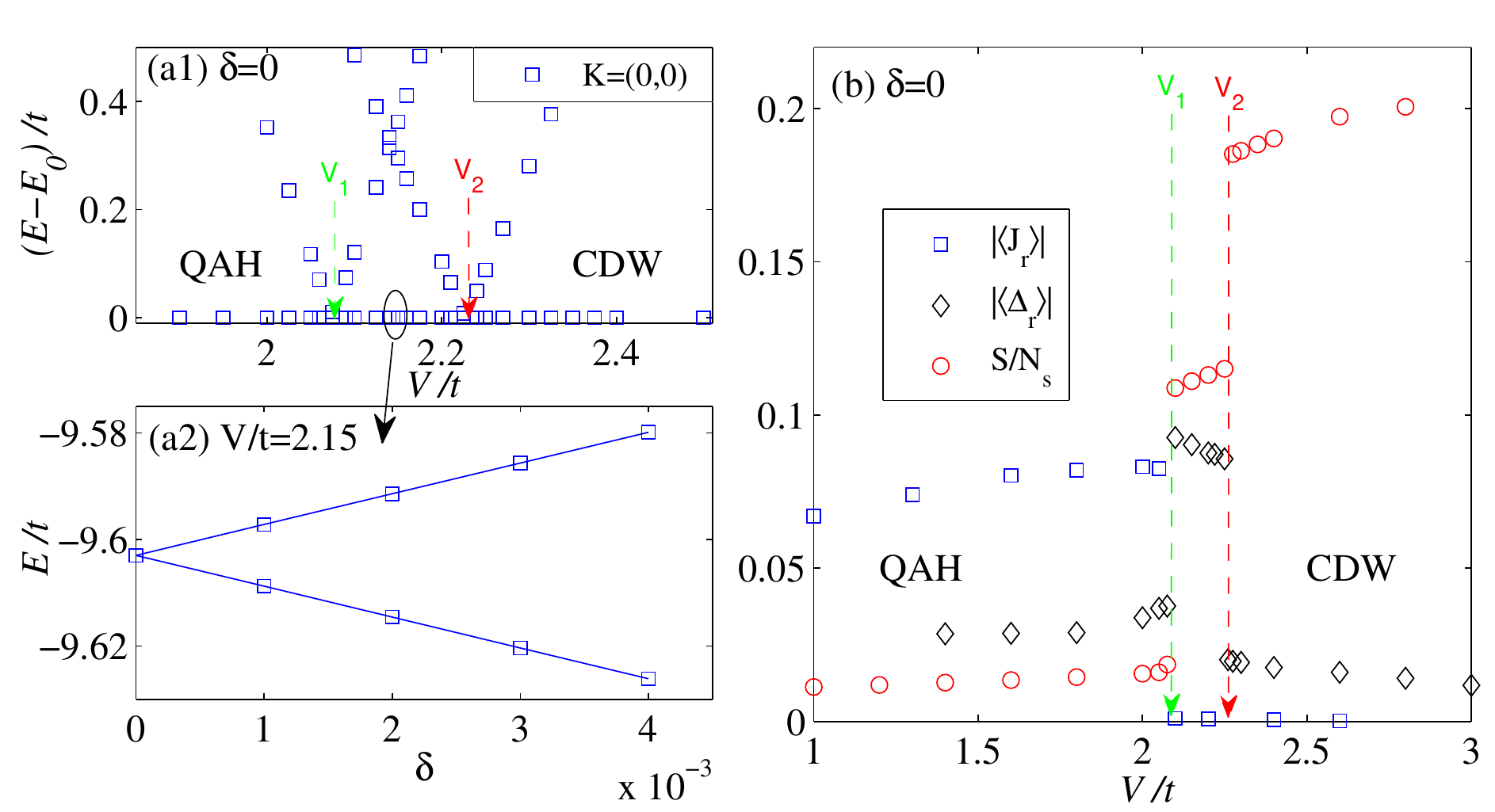}
  \caption{\label{phaseED}(Color online) Numerical ED results for spinless fermions at half-filling $N_s=2\times4\times4=32,N=16$ in the checkerboard lattice with $V=t,t'=0.8t,\delta=m=0$. (a1) The evolutions of the low energy spectrum as a function of $V$. Only the lowest five energy levels are shown. (a2) The perturbation response induced by $\delta$. (b) The evolutions of  $\langle J_{\rr}\rangle,\langle\Delta_{\rr}\rangle,S/N_s,\Delta_c/4$  as a function of $V$.}
\end{figure}

\textit{Intermediate Phase.---}
We turn to analyze the emergent intermediate phase between the QAH phase and the nematic CDW phase. Figure~\ref{phaseED}(a1) depicts the evolution of the interacting many-body low energy spectrum.
Near the point $V=V_1$, the doublet ground states of the QAH phase undergo a level crossing with the two-fold degenerate excited levels.
When the interaction $V$ increases further, there appears another level crossing near the transition around  $V=V_2$.
For $V_1<V<V_2$, the doublet ground states host  nonzero bond orders $\Delta_{\rr}\neq0$, which is calculated from the symmetry-breaking perturbation response  $\Delta_{\rr}=\frac{1}{2N_s}\frac{\partial E(\delta)}{\partial\delta}|_{\delta=0}=|a_{\rr}^{\dag}a_{\rr+\widehat{e}_x}|-|a_{\rr}^{\dag}a_{\rr+\widehat{e}_y}|
=|b_{\rr}^{\dag}b_{\rr+\widehat{e}_x}|-|b_{\rr}^{\dag}b_{\rr+\widehat{e}_y}|$ ~\cite{note}. For example,  we have $\Delta_{\rr}\sim 0.08$ at $V=2.15$, as shown in Fig.~\ref{phaseED}(a2). The nonzero bond order indicates the bond nematic nature of the intermediate phase.

Meanwhile, we plot the evolutions of the current order parameter $\langle J_{\rr}\rangle$,
sublattice bond order parameter $\langle\Delta_{\rr}\rangle$, nematic CDW density structure factor $S/N_s$ and the charge-hole gap $\Delta_c=(E_0(N+1)+ E_0(N-1)-2E_0(N))/2$ as a function of $V$ in Fig.~\ref{phaseED}(b).
For weak interactions $V<V_1$, $\langle J_{\rr}\rangle$ has  a finite expectation value, and $\Delta_c$ increases with the increase of the  interaction strength, manifesting the robustness of a gapped QAH phase.
On the other hand,  both $\langle\Delta_{\rr}\rangle$ and $S/N_s$  take small values consistent with the properties of a gapped  QAH phase.
When interaction $V$ goes across $V_1$,  $\langle J_{\rr}\rangle$,  $\langle\Delta_{\rr}\rangle$, and $S/N_s$ experience a sudden jump, where $\langle J_{\rr}\rangle$ drops down to a vanishingly small value of the order $10^{-4}$, and $\Delta_c$ begins to decrease quickly, signaling the collapsing of a topological phase
(finite size scaling results  for the $\Delta_c$ will be discussed below).
Simultaneously,  $\langle\Delta_{\rr}\rangle$, and
$S/N_s$ jump  to a finite large value.  When interaction further increases beyond the critical value $V_2$, $\langle\Delta_{\rr}\rangle$ drops down to a vanishingly small value of the order $10^{-5}$ while
$S/N_s$ undergoes another step jump to a  larger value, where the system becomes  an insulating nematic CDW phase with a large excitation gap as shown in Fig.~\ref{phaseED}(a1).

\begin{figure}[t]
  \includegraphics[height=1.65in,width=3.4in]{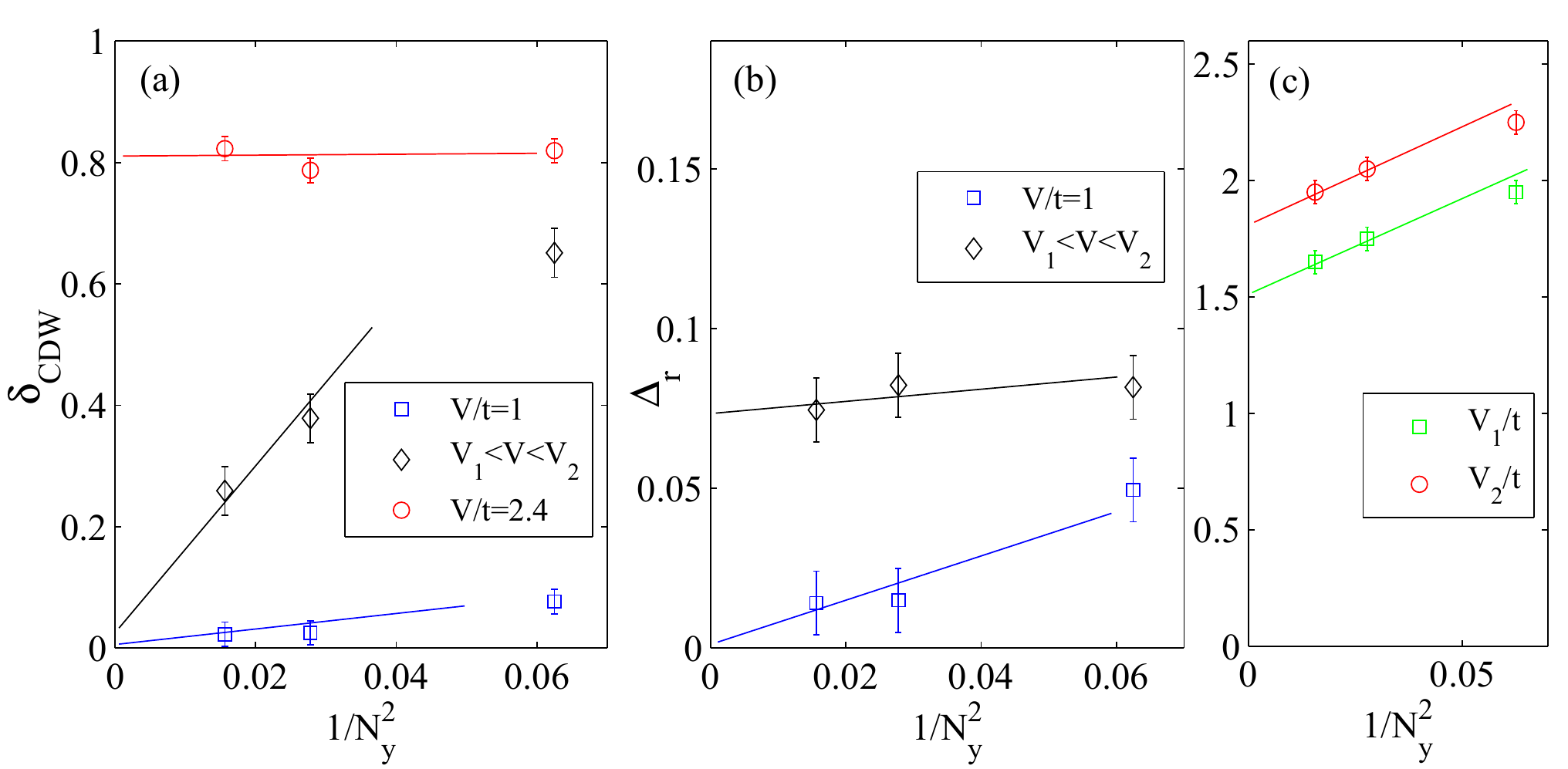}
  \caption{\label{scale}(Color online) Finite-size scaling of DMRG results for (a) nematic CDW order, (b) sublattice bond order and (c) transition point.
  All the parameters are the same as those in Figs.~\ref{dmrgc}(a-c).
 }
\end{figure}

\begin{figure}[t]
  \includegraphics[height=1.4in,width=3.4in]{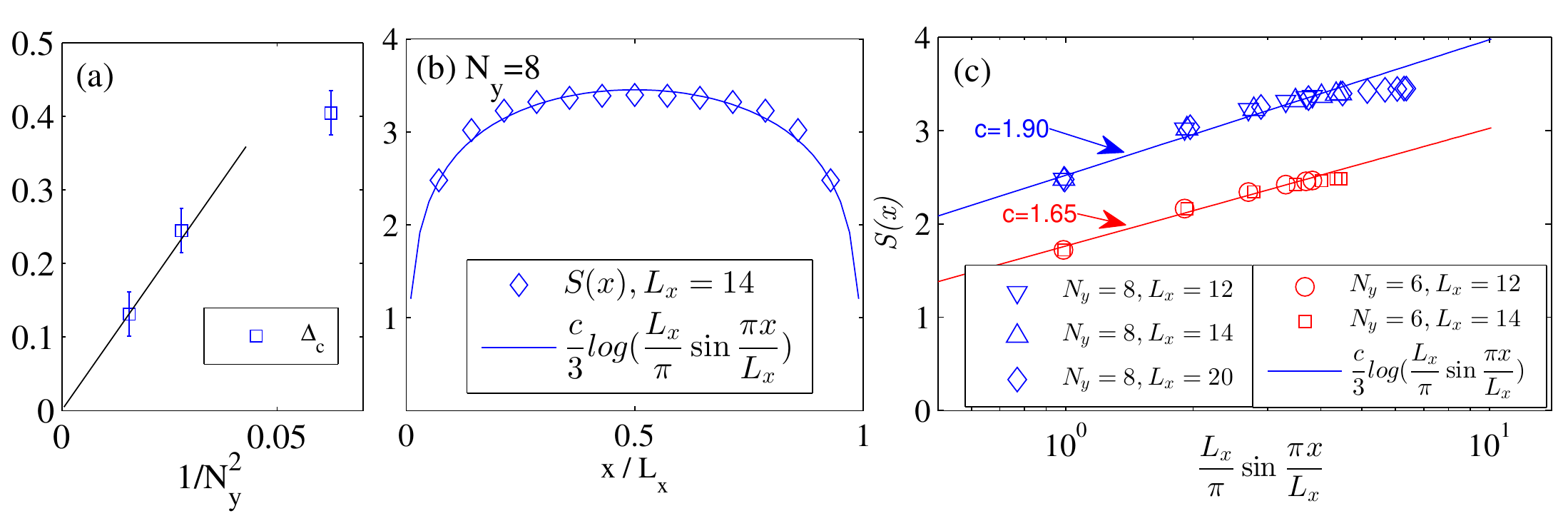}
  \caption{\label{entropy}(Color online) DMRG results in the intermediate region $V_1<V<V_2$ for (a) finite-size scaling of the charge-hole gap $\Delta_c$; (b) the von Neumann entanglement entropy $S(x)$ for a given system size; (c) finite-size logarithmic scaling of entanglement entropy $S(x)$ with different cylinder widths $N_y$ and lengths $L_x$, in comparison with the theoretical prediction $S(x)=\frac{c}{3}\log(\frac{L_x}{\pi}\sin\frac{\pi x}{L_x})$.
  All the parameters are the same as those in Fig.~\ref{scale}.
 }
\end{figure}

As presented  above,  based on a given system size calculation, we see the intermediate phase between $V_1$ and $V_2$ is associated with a finite bond order $\langle\Delta_{\rr}\rangle$.
To inspect properties of the phase in the thermodynamic limit,
we carry out a finite size scaling of bond order $\Delta_{\rr}$ and nematic order $\delta_{CDW}=\langle n_{\rr}^{a}-n_{\rr'}^{b}\rangle=\frac{2}{N_s}\frac{\partial E(m)}{\partial m}$. In Fig.~\ref{scale},
it is found that $\Delta_{\rr}$ extrapolates to  a finite value, but $\delta_{CDW}$ gradually decreases down to a negligibly small value in the limit $1/N_y^2 \rightarrow 0$. Thus,
our ED and DMRG methods provide  a very strong evidence for the existence of the intermediate topologically trivial bond-ordered phase.
Physically, the doublet states of the QAH phase host opposite eigenvalues $\pm 1$ angular momentum
in the presence of $C_4$ rotation symmetry and time-reversal symmetry,
while both bond-ordered and nematic CDW phase breaks $C_4$ symmetry down to $C_2$ symmetry. In Ref.~\cite{Wu2016},
such an argument is used to claim a first-order phase transition between $C_4$ symmetric phases and $C_2$ symmetric phases.
Our ED and DMRG study access  much larger systems, which lead to the discovery of a time-reversal symmetric intermediate bond-ordered phase
sandwiched in between QAH and CDW phases~\cite{notex}.  At the mean-field level \cite{Sun2009}, one can show that this phase is gapless with the QBT splitting into two Dirac
points (see Fig. \ref{phase}). Moreover,
we further provide strong  numerical evidences to support the gapless nature of the intermediate phase, which point to the existence of Dirac cone structure.
First, we find that the finite-size scaling of the charge-hole gap $\Delta_c$ in Fig.~\ref{entropy}(a) gives a nearly zero value in the thermodynamic limit, which
implies the gapless single-particle excitation nature.
Second, the entropy dependence on the cylinder length approaches an arch structure as indicated in Fig.~\ref{entropy}(b), and it can be fitted to  the universal scaling function (up to an additive constant depending on the cylinder width)~\cite{Sheng2009,Ju2012,Chen2015} $S(x)=\frac{c}{3}\log(\frac{L_x}{\pi}\sin\frac{\pi x}{L_x})$ with the central charge $c\approx 2$, as plotted in Fig.~\ref{entropy}(c).
The only deviation from the straight line fitting  shown in Fig. 5(c) appears at larger $x$ values due to the convergence
difficulty in capturing the entanglement of a gapless system.  
The finite gap for finite size systems shown in Fig. 5(a) can be understood as the following.
On the finite  size systems, the available momentum points in the Brillouin zone are discrete, such that the Dirac points are not guaranteed to be exactly covered in our calculations and a finite-size gap appears. 
Similarly, when experienced a small energy gap, the central charge obtained from entropy behavior would deviate from the ideal value $c=2$ of free Dirac fermions. 
Nevertheless, the finite-size scaling of the gap,  and  the entropy scaling approaches the Dirac liquid behavior  in the thermodynamic limit,  as indicated  in Figs.~\ref{entropy}(a) and~\ref{entropy}(c).  These  numerical results indeed support that both the charge-hole gap $\Delta_c\rightarrow0$ and the central charge $c\rightarrow2$ as the cylinder width $L_y=2N_y$ increases.

\textit{Phase Transition.---}
To further study phase transitions, we exploit the finite DMRG calculation on a cylindrical geometry up to a maximum width $L_y=16$ ($N_y=8$) and length $L_x=N_x=20$.
We measure five different physical quantities of the ground state as a function of $V$:
the wavefunction overlap $F(V)=|\langle\psi(V)|\psi(V+\delta V)\rangle|$ ($\delta V$ is as small as $0.1t$),
the entanglement entropy $S_L$ in the middle of the cylinder, in addition to order parameters $\Delta_{\rr}$,
$\delta_{CDW}$ and $J_{\rr}$. The first-order transition is characterized by the discontinuous behavior of these physical quantities.

\begin{figure}[t]
  \includegraphics[height=3.0in,width=3.3in]{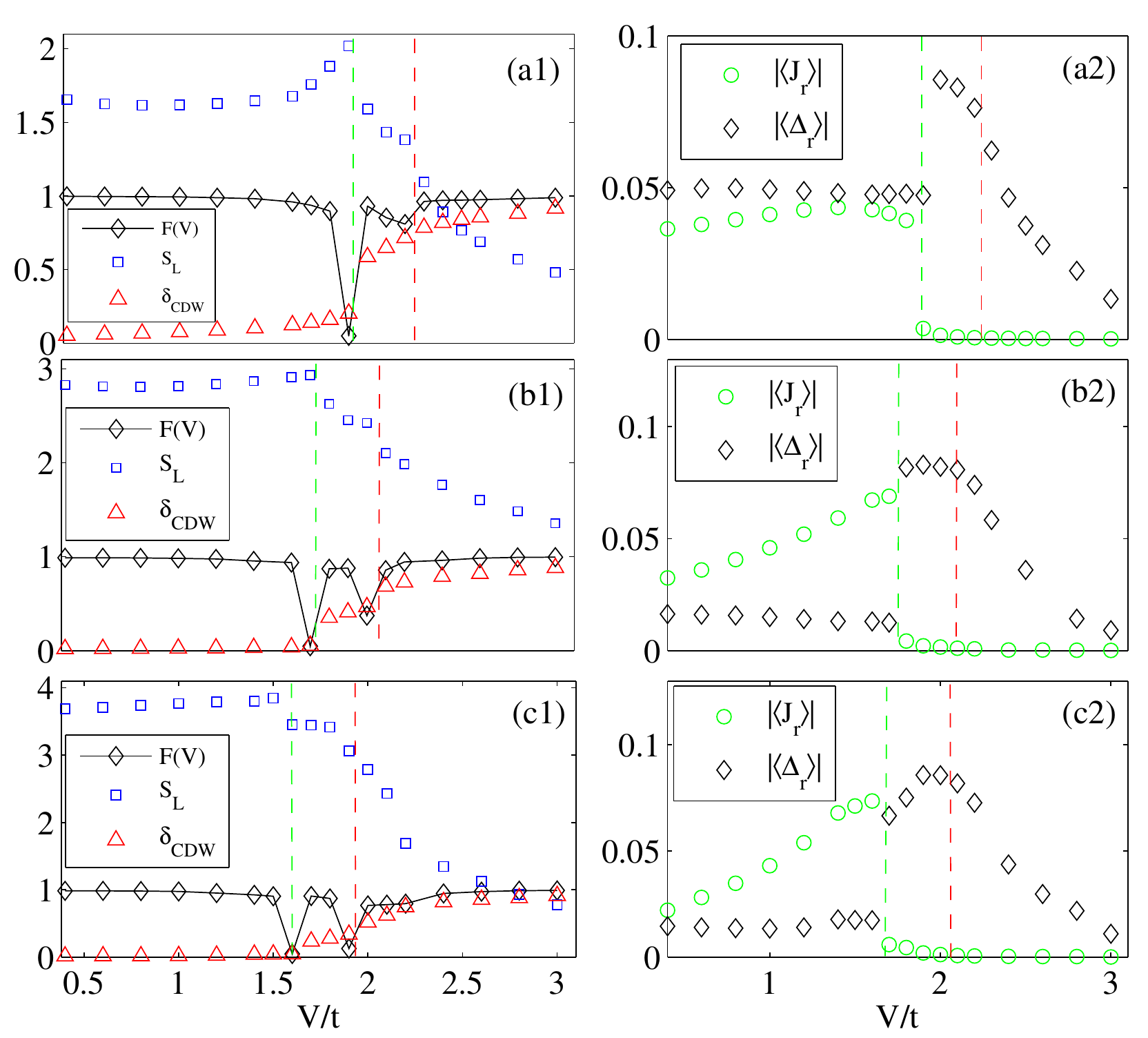}
  \caption{\label{dmrgc}(Color online) Numerical finite DMRG results in a cylinder checkerboard lattice with width length $L_x=N_x=20$ at half-filling.
  The evolutions of $F(V)=|\langle\psi(V)|\psi(V+\delta V)\rangle|$, $S_{L}$, $\delta_{CDW}$, $\Delta_{\rr}$ and $\langle J_{\rr}\rangle$ as a function of $V$ for different cylinder widths (a) $N_y=4,L_y=8$,
  (b) $N_y=6,L_y=12$ and (c) $N_y=8,L_y=16$. The dashed lines indicate the transition points. The parameters $t'=0.8t,\delta=m=0$, and the maximally kept number of states 3620.}
\end{figure}

As shown in Figs.~\ref{dmrgc}(a-c), for $V<V_1$, $F(V)$ has a large value close to 1,
 both $S_L$ and $\delta_{CDW}$ exhibit featureless properties,
but $J_{\rr}$  grows slightly with the increase of $V$, implying the robustness of the QAH phase.
When $V$ approaches a transition point $V_1$, $F(V)$ suddenly drops down to a very small value close to zero,
and both order parameters $\Delta_{\rr},J_{\rr}$ exhibit a sharp discontinuous jump near the transition point.
Similarly, 
$S_L$ starts to drop at $V\simeq V_1$, with a discontinuous derivative $\partial S_L/\partial V$. These
results are consistent with a first order transition between the QAH and intermediate phases~\cite{Zozulya2009,Zeng2017}.
When the interaction $V$ increases further close to $V_2$, $\delta_{CDW}$ undergoes a  jump to a large saturated value close to 1,
while $\Delta_{\rr}$ gradually drops down to zero in the strongly interacting regime. $F(V)$ exhibits a minimum, sharpening as the cylinder width increases from $N_y=4$ to $N_y=8$,
while  entanglement entropy $S_L$ also drops down to a smaller value consistent with an insulating phase at $V>V_2$.
These results are consistent with  a first-order  transition into a CDW phase at $V=V_2$.

\section*{Discussions}
In summary, we have numerically presented a solid diagnosis of 
an interaction-driven spontaneous QAH phase in the checkerboard lattice with a quadratic band touching by turning on any weak interaction.
Such a diagnosis relies on finite size scaling up to wide systems ($L_y=20$ lattice spacing)  and use detecting flux  for the QAH.
The QAH phase hosts two-fold ground state degeneracies with opposite spontaneous TRS breaking behaviors,
and a quantized Hall conductance measured by Laughlin argument of charge pumping.
In particular, we demonstrate the existence of a bond-ordered phase sandwiched between the QAH phase and the nematic CDW phase, characterized by the sublattice bond order. 
 The intermediate bond-ordered phase points to the long-sought Dirac liquid phase in the QBT systems. \cite{Sun2009,Pujari2016}
We believe that this work would open a new route for  the study of the possible competing intermediate phases under the interplay of interaction and frustration, and excite a more extensive investigation of the fate of the QAH phases in many other systems, such as bilayer graphene~\cite{Nandkishore2010,Pujari2016}, and $C_6$ symmetric Kagome lattice~\cite{Wen2010} where an intermediate gapless CDW phase was also proposed.
Other future directions include a study of the interplay of nearest-neighboring and next-nearest-neighboring interactions,
which may lead to rich possibility and other competing phases driven by interactions~\cite{Sur2018}.
At experimental side,  our work also suggests a practical way of opening and closing the QBT gap, inducing quantized
dissipationless transport currents by interactions.

\section*{Methods}
\textit{Exact diagonalization}\\
We  perform  ED  calculations  on  the  model Eq. (1) with parameters
as indicated in the corresponding text and figure captions.
In the ED calculations, we study the many-body ground state of $H$ at half-filling in a finite system of $N_x\times N_y$ unit cells (the total number of sites $N_s=2\times N_x\times N_y$).
The energy eigenstates  are labeled by a total momentum $K=(K_x,K_y)$ in units of $(2\pi/N_x,2\pi/N_y)$ in the Brillouin zone.

\textit{Density-matrix renormalization group}\\
For larger systems we exploit both finite and infinite DMRG on the cylindrical geometry. We keep the dimension of DMRG kept
states up to 5600 to obtain accurate results (the truncation error is of the order $10^{-6}$).
This leads to excellent convergence for the results that we report here.
The geometry of cylinders is open boundary condition in the $x$-direction and periodic boundary condition in the $y$-direction.

\vspace{10pt}

\textit{Note add.---} In the preparation of this work, we become aware of a parallel work from Ref.~\cite{Sur2018}.

\textit{Acknowledgement}---T.S.Z thanks Chuanwei Zhang for support and encouragement on studying quadratic band touching.
We thank S. S. Gong for private communication prior to publication. T.S.Z acknowledges the support from Air Force Office of Scientific Research (FA9550-16-1-0387), National Science Foundation (PHY-1505496),
and Army Research Office (W911NF-17-1-0128).
W. Z. was supported by Department of Energy (DOE) National Nuclear Security Administration through Los Alamos National Laboratory LDRD Program.
D.N.S.  was supported by  the DOE, through  the Office of Basic Energy Sciences under the grant No. DE-FG02-06ER46305.

\textit{Author contributions.---}
W.Z. and D.N.S. proposed the idea. T.S.Z. proposed the model and carried out the calculation and drafted the article.
All authors contribute to the writing and revision of the manuscript.

\textit{Competing interests.---} The authors declare no competing interests.

\textit{Data Availability---} The data that support the findings of this study are available from the corresponding author upon reasonable request.


\begin{thebibliography}{99}
\bibitem{note} The bonds $a_{\rr}^{\dag}a_{\rr+\widehat{e}_x}$ and $a_{\rr}^{\dag}a_{\rr+\widehat{e}_y}$ have opposite signs and different magnitudes. Also, $a_{\rr}^{\dag}a_{\rr+\widehat{e}_x}=-b_{\rr}^{\dag}b_{\rr+\widehat{e}_x}$ and $a_{\rr}^{\dag}a_{\rr+\widehat{e}_y}=-b_{\rr}^{\dag}b_{\rr+\widehat{e}_y}$.
\end{thebibliography}
\end{document}